# Leveraging Mobile Learning Platforms for Flexible Education Delivery: Bridging Educational Gaps in Afghanistan

Mursal Dawodi
mursal.dawodi@gmail.com
FemTech

Jawid Ahmad Baktash
jawid.baktash1989@gmail.com
FemTech

Sayed Mohammad Reza Dawodi
sayed.moh.reza@gmail.com
FemTech

**Abstract**
The educational landscape of Afghanistan, besieged by infrastructural inadequacies and socio-political tribulations, presents a compelling case for the integration of mobile learning platforms. This article embarks on an exploratory voyage into the realms of mobile learning as a potential harbinger of educational transformation in Afghanistan. It delineates the pervasive educational challenges, underscores the technological innovations powering mobile learning platforms, and illuminates the pathways through which mobile learning can transcend the extant barriers to education. Enriched by real-world case studies, the narrative unravels the pragmatic lessons that can be harnessed to tailor mobile learning solutions to Afghanistan's unique context. The discussion further traverses the collaborative horizon, elucidating the synergistic interplay among academia, government, the private sector, and international bodies essential for the successful implementation of mobile learning platforms. The article also furnishes pragmatic recommendations, emphasizing the triad of policy formulation, infrastructure enhancement, and capacity building as cornerstone imperatives. The envisioned integration of mobile learning platforms augurs a paradigmatic shift towards a more accessible, inclusive, and resilient educational framework in Afghanistan, with far-reaching implications for socio-economic development. Through a meticulous amalgamation of technology, policy, and collaborative endeavors, this article posits that Afghanistan stands on the cusp of an educational renaissance, with mobile learning platforms serving as a pivotal conduit towards this envisioned horizon.

## 1 Introduction

Afghanistan, with a historical tapestry rich in tradition and culture, has been beleaguered by decades of conflict, political turmoil, and social unrest. These adversities have left indelible imprints on the nation's educational infrastructure. Over the years, the country has grappled with an array of educational challenges that have hampered the progress of human capital development.

Historically, the educational scenario in Afghanistan has been notably gender-biased, with women and girls being systematically denied access to education. The Soviet invasion in 1979 and the subsequent civil war further exacerbated the education crisis, leading to severe destruction of educational infrastructure, massive displacement of populations, and a devastating loss of human lives (Bureau of South and Central Asian Affairs, 2011).

The rise of the Taliban regime in the 1990s imposed stringent restrictions on education, especially for females, further dwindling the educational prospects in the country (Razia, 2009). Post 2001, with the fall of the Taliban regime, Afghanistan witnessed a resurgence of hope for its educational sector. International aid and the commitment of the Afghan government led to the reopening of schools and universities, the training of teachers, and the development of curricula (UNICEF, 2011).

However, despite these positive strides, the progress remained fragile and was continually threatened by ongoing conflict, poverty, cultural norms, and geographical barriers. With the re-emergence of the Taliban in 2021, the future of education, particularly for females, hangs in the balance. The ban on girls' and women's access to higher education and work since December 2022 underscores the regressive educational policies that threaten to undo the progress made over the last two decades (UNESCO, 2022; UNICEF, 2022).

Amidst this grim backdrop, global initiatives aimed at promoting education in crisis regions have emerged as a beacon of hope. Organizations like UNESCO, UNICEF, and the World Bank have been actively involved in spearheading educational projects in crisis-hit regions. For instance, the Global Partnership for Education (GPE) has been instrumental in supporting education in conflict-affected countries, ensuring that children have access to quality education even in the most challenging circumstances (GPE, 2019).

Afghanistan stands to benefit immensely from such global initiatives. The infusion of international aid, technical expertise, and the adoption of innovative educational solutions like Mobile Learning (M-learning) can play a pivotal role in revamping the educational landscape of the country. By harnessing the potential of M-learning, Afghanistan can overcome geographical and infrastructural barriers, promote gender inclusivity in education, and foster a conducive learning environment for all.

The ensuing sections will delve deeper into the technological advancements powering Mobile Learning Platforms, real-world case studies demonstrating their impact, and the vision for integrating such platforms within Afghanistan's educational framework to catalyze broader socio-economic development.

## Mobile Learning: A Beacon of Hope

Mobile Learning (M-learning) emerges as a viable solution in this exigent scenario. The global mobile learning market burgeoned from $53.94 billion in 2022 to $69.33 billion in 2023, with a notable CAGR of 28.5% (Statista, 2023). M-learning provides an "anytime, anywhere" learning experience, enabling access to educational resources irrespective of geographic or infrastructural barriers (Ally, 2009). This is particularly pertinent in developing countries, including Afghanistan, where the digital divide and lack of educational infrastructure are pronounced (UNESCO, 2017).

The rapid development of digital learning platforms globally, prompted by school closures during the COVID-19 pandemic, underscores the potential of technology in enhancing educational access and quality beyond traditional classroom settings (Huang et al., 2020). With the convergence of web functionality onto mobile platforms, coupled with the affordability and portability of communication devices, m-learning is poised to revolutionize education delivery (Kukulska-Hulme, 2012).

# 2 Literature Review

Mobile learning (M-learning) heralds a significant paradigm shift in the field of educational technology, offering a modern approach to knowledge acquisition and dissemination beyond traditional classroom confines. Over the last decade, numerous studies and theories have been propounded, elucidating the potential of mobile learning platforms in bridging educational gaps, especially in crisis regions. This literature review endeavors to encapsulate key findings from various seminal and contemporary studies, while exploring different educational models pertaining to mobile learning.

## 2.1 Theories Underpinning Mobile Learning

Various theories underscore the conceptual framework of mobile learning, shedding light on diverse aspects of this modern educational paradigm. Among these, Siemens' (2005) theory of Connectivism emerges as a cornerstone, articulating a learning model for the digital era. It underscores the significance of networks, the diversity of opinions, and the indispensable role of technology in cultivating an environment conducive to continuous learning. This theory essentially mirrors the digital lattice within which mobile learning operates, fostering a milieu of interconnected knowledge spheres.

In a similar vein, the theory of Constructionism, as postulated by Papert (1991), dovetails with the ethos of mobile learning. It encapsulates the essence of learning-by-doing, a premise that mobile learning platforms ardently uphold. By facilitating a digital arena where learners can interact, explore, and construct knowledge, mobile learning platforms embody the crux of Constructionism. They provide a digital scaffold where learners can actively engage with content, thereby fostering a deeper understanding and retention of knowledge.

Further, the realm of mobile learning significantly resonates with Knowles' (1975) concept of Self-Directed Learning. This concept accentuates the empowerment of individuals in steering their learning journey—controlling the pace, trajectory, and assessment of their learning endeavors. Mobile learning platforms epitomize this empowerment by offering a flexible and personalized learning experience. They allow learners to tailor their learning pathway, adjust the pace of learning, and self-assess their progress. Through this lens, mobile learning unfolds as a medium that not only delivers content but also instills a sense of agency and self-direction in learners, nurturing a more engaging and personalized learning experience.

## 2.2 Impact in Crisis Regions

Several studies have shed light on the transformative potential of mobile learning, particularly in crisis-stricken regions, showcasing its capability to mitigate educational inequities. One of the significant advantages of mobile learning highlighted in literature is its ability to ensure educational accessibility and continuity amidst adversities. For instance, a pivotal study conducted by UNESCO in 2012 accentuated the critical role mobile learning platforms play in sustaining educational endeavors in regions beleaguered by crises, where conventional schooling often faces abrupt disruptions (UNESCO, 2012). This aspect of mobile learning is particularly pertinent in contexts where political turmoil or natural disasters recurrently thwart the normalcy of educational institutions, thus, underscoring the imperative of alternative learning avenues.

Furthermore, the domain of gender inclusivity, which is a cornerstone for equitable education, has also been positively impacted by mobile learning platforms. In conservative societies where socio-cultural norms often pose barriers to women and girls' education, mobile learning emerges as a beacon of hope. A study cited by Vosloo in 2013 delineates how mobile learning platforms have been instrumental in fostering gender inclusivity by providing women and girls with the means to access education remotely (Vosloo, 2013). This aspect of mobile learning is not only pivotal for achieving gender parity in education but also for promoting social inclusivity and empowerment. Through the lens of these studies, it's palpable that mobile learning is not merely a technological advancement, but a vehicle for social transformation, addressing the educational chasms exacerbated by crises and conservative societal norms.

## 2.3 Global Utilization of Mobile Learning Platforms

Various educational models have been employed globally to integrate mobile learning platforms into mainstream education, each with its unique approach and benefits. One such model is the Flipped Classroom Model, which leverages mobile learning platforms to deliver content outside the classroom. This approach frees up classroom time for interactive and collaborative learning activities, enhancing the learning experience by allowing for more hands-on and interactive sessions (Bishop & Verleger, 2013). On the other hand, the Blended Learning Model integrates mobile learning with traditional classroom instruction to provide a more flexible and personalized learning experience. This model aims to balance the benefits of face-to-face instruction with the flexibility and personalized learning opportunities offered by mobile learning platforms (Driscoll, 2002). Lastly, the Self-Paced Learning Model is centered on allowing learners to progress through the curriculum at their own pace, facilitated by the flexibility of mobile learning platforms. This model caters to individual learning speeds and preferences, making learning more learner-centric and adaptable to individual needs.

## 2.4 Real-World Implementations

Mobile learning platforms have garnered traction across a plethora of geographical and socio-political landscapes, each presenting a unique set of challenges and outcomes. For instance, in Kenya, the advent of Shupavu291, a text-message-based mobile learning platform, has played a pivotal role in ensuring the continuity of education amidst socio-political disruptions (West, 2017). This initiative has demonstrated the resilience and adaptability inherent in mobile learning platforms to provide educational resources in challenging circumstances.

Across the continent, in Pakistan, a study exploring the readiness of university teachers and students towards mobile learning unveiled a substantial level of readiness among educators to embrace mobile learning tools (Shah et al., 2019). This finding elucidates a conducive environment for the integration and acceptance of mobile learning solutions within the higher education sector. The readiness among educators, a crucial stakeholder group, bodes well for the successful deployment and uptake of mobile learning platforms.

Furthermore, the initiative by Colégio Miguel de Cervantes in Brazil, showcased as part of UNESCO's compilation of best practices in mobile learning, has empowered students to morph into agents of social transformation through mobile learning (UNESCO, 2017). This initiative underscores the transformative potential of mobile learning not only in fostering educational engagement but also in catalyzing social change. The case of Brazil accentuates the broader societal impact that can be engendered through well-designed and effectively implemented mobile learning initiatives.

These diverse implementations of mobile learning platforms across Kenya, Pakistan, and Brazil underscore the adaptability and transformative potential inherent in mobile learning solutions. They provide a rich tapestry of experiences and outcomes that can inform and inspire the deployment of mobile learning platforms in other similar contexts, such as Afghanistan. Through a judicious adaptation of mobile learning platforms to local contexts, it is plausible to envisage a substantial positive impact on education delivery and social transformation, transcending geographical and socio-political barriers.

# 3 Methodology

The methodology section delineates the approach adopted to gather and analyze data pertinent to the exploration of mobile learning platforms as a viable solution for bridging educational gaps in Afghanistan. Given the nascent stage of mobile learning implementation in Afghanistan, this study predominantly relies on a comprehensive review of existing literature, case studies, and secondary data from reputable sources. The methodology encompasses the following steps:

- Literature Review:
    A thorough review of existing literature was conducted to glean insights into theories underpinning mobile learning, its global implementations, and its potential impact in crisis regions. Peer-reviewed articles, reports from educational organizations, and publications from global educational initiatives were perused to construct a well-rounded understanding of the subject matter.
- Surveys and Interviews:
    While primary data collection through surveys and interviews was envisaged, the prevailing socio-political conditions in Afghanistan posed significant challenges to this endeavor. Hence, this study leaned more towards an exhaustive review of secondary data.
- Case Studies Analysis:

An examination of real-world implementations of mobile learning platforms in other developing countries was undertaken to garner practical insights. This analysis aimed to understand the strategies employed, challenges encountered, and the outcomes achieved through mobile learning initiatives.

- Comparative Analysis:

A comparative analysis was performed to juxtapose the mobile learning implementations in different socio-political and geographical contexts against the unique challenges posed by the Afghan educational landscape. This analysis sought to derive actionable insights that could inform the effective integration of mobile learning platforms in Afghanistan.

# 4 Technological Innovations in Mobile Learning

The burgeoning realm of mobile learning (M-learning) is intricately tethered to swift advancements across various technological spectrums. A notable stride is the synthesis of Artificial Intelligence (AI), Augmented Reality (AR), Virtual Reality (VR), and Cloud Computing with mobile learning platforms, a fusion that has considerably broadened the horizons and efficacy of educational delivery channels. This segment delves into these groundbreaking technological innovations, exploring their potential customization to address the distinct needs and hurdles inherent in the Afghan educational landscape.

Embarking on the journey with Artificial Intelligence (AI), it stands as a linchpin in fostering personalized learning trajectories. Through meticulous analysis of learners' data, AI unravels insightful patterns, strengths, and areas necessitating enhancement. The integration of AI within mobile learning platforms paves the way for delivering bespoke content, recommending pertinent resources, and imparting tailored feedback, thus significantly enriching the learning expedition. Tailoring this to the Afghan scenario, the realms where teacher proficiency and resources are often found wanting, AI-driven mobile learning platforms hold the promise of substantially augmenting both teaching and learning paradigms. By carving out personalized learning routes, AI harbors the potential to mend the educational fissures prevalent in the nation.

Transitioning to Augmented Reality (AR) and Virtual Reality (VR), these technologies usher learners into immersive educational realms, rendering the learning experience more captivating and interactive. While AR seamlessly superimposes digital data onto the tangible world, VR transports learners into a wholly immersive digital milieu. Adapting this to the Afghan context, the infrastructural impediments can be adeptly navigated through AR and VR by granting students virtual access to laboratories, excursions, and interactive 3D models. This not only compensates for the dearth of physical resources but also unlocks doors to experiential learning within a secure, regimented ambiance.

Lastly, the advent of Cloud Computing emerges as a beacon of hope in transcending geographical and infrastructural barricades, by enabling the storage and retrieval of educational resources anytime, anywhere. When tailored to the Afghan setting, cloud computing can centralize educational resources, making them accessible to both students and educators across the expanse of Afghanistan, irrespective of their geographical bearings. This holds particular resonance in remote or conflict-afflicted zones where educational infrastructure either scantily exists or is entirely absent. Through cloud computing, the realms of learning can be made boundlessly accessible, embodying a substantial stride towards bridging the educational chasms that beleaguer Afghanistan.

## 4.1 Implications for the Afghan Education System

The integration of technological advancements with mobile learning platforms holds a promising potential for revolutionizing the education system in Afghanistan. Through optimizing scarce educational resources, technologies like Artificial Intelligence (AI) and Cloud Computing centralize educational content and facilitate remote access, making education more accessible. Augmented Reality (AR) and Virtual Reality (VR) stand at the forefront of enhancing student engagement and retention by delivering interactive and immersive learning experiences. Moreover, AI extends its utility to the realm of teacher training and professional development by unlocking access to a wealth of educational resources and global best practices. In the conflict-afflicted terrains of Afghanistan, these technologies can underpin mobile learning platforms to ensure continued education in a safe, virtual environment. Furthermore, the scalability bestowed by Cloud Computing enables the expansion of mobile learning platforms to accommodate a wider student base without necessitating substantial infrastructural investments. This blend of AI, AR, VR, and Cloud Computing with mobile learning platforms unveils a spectrum of opportunities tailored to address the unique challenges faced by the Afghan education system. It paves the way towards bridging the prevailing educational gaps and cultivating a conducive learning environment, notwithstanding the challenging circumstances endemic to the region. Through a judicious harnessing of these technologies, Afghanistan stands on the cusp of an educational transformation that could significantly ameliorate the educational delivery and outcomes amidst its prevailing adversities.

# 5 Real-world Case Studies

The potential of mobile learning platforms to bridge educational gaps is vividly illustrated through various real-world case studies across differing contexts. A notable example comes from Kenya, where during times of political unrest and social disruption, particularly during the 2017 presidential election, a text-message-based mobile learning platform named Shupavu291 proved instrumental. Students utilized this platform to sustain their educational endeavors amidst widespread school closures. This scenario accentuates how mobile learning platforms can provide a resilient educational infrastructure during times of social upheaval.

Broadening the scope to developing countries at large, the advent of mobile technologies presents a promising avenue to deliver education without the dependency on extensive traditional communications infrastructure. This technological leapfrogging could circumvent certain developmental phases typical in developed countries, such as the establishment of extensive electricity power grids and the construction of multiple computer rooms (Kizilcec et al., 2021). The flexibility and accessibility of mobile learning platforms can substantially mitigate infrastructural constraints, thus bridging educational gaps in resource-limited settings.

On a global scale, initiatives like UNESCO's 5-year project launched in 2016 underscore the concerted efforts to harness mobile learning for creating inclusive and equitable learning environments. This project aimed to guide school-wide planning and utilization of mobile learning to ensure the quality, effectiveness, and relevance of education in a digital world (Traxler & Kukulska-Julme, 2005). Such initiatives elucidate the global recognition and concerted efforts towards leveraging mobile learning as a vehicle for educational inclusivity and quality.

Venturing into Nigeria, the development of a mobile learning application named MobileEdu for higher education stands as a testament to the adaptability of mobile learning platforms. This application, designed to facilitate the learning of computer science courses on mobile devices, embodies the ubiquitous, collaborative, and social aspects of learning among higher education students (UNESCO, 2023). The case of Nigeria exemplifies how mobile learning platforms can be tailored to cater to the specific educational needs and contexts, thereby enhancing the learning experience and accessibility for students.

Collectively, these diverse case studies from Kenya, developing countries at large, global initiatives, and Nigeria, furnish compelling evidence on the effectiveness and adaptability of mobile learning platforms in bridging educational gaps. They provide a rich repository of practical insights that could be instrumental in guiding the implementation of mobile learning platforms in other contexts, including Afghanistan, to foster a more inclusive and resilient educational framework.

### 5.1 Lessons Learned and Implications for Afghanistan

The exploration into mobile learning platforms reveals multifaceted benefits that can be harnessed to ameliorate educational constraints, particularly in crisis-prone regions like Afghanistan. A quintessential example is the case of Shupavu291 in Kenya, which showcases the vital role of mobile learning platforms in ensuring education continuity amid crises. This scenario is markedly pertinent for Afghanistan, where the socio-political milieu often disrupts traditional educational delivery mechanisms. Moreover, the feature of infrastructure independence inherent in mobile learning platforms emerges as a significant boon. These platforms possess the capability to circumvent infrastructural hurdles, a lesson invaluable for Afghanistan where the educational infrastructure frequently ranges from lacking to dilapidated.

On a broader spectrum, global collaboration surfaces as a pivotal element in leveraging mobile learning for educational equity. The UNESCO project accentuates this notion, advocating for a global collaborative approach towards mobile learning integration. This perspective could be notably embraced by Afghanistan to fortify its educational framework, drawing upon international expertise and resources.

Furthermore, the emphasis on localized solutions is underscored by the case of MobileEdu in Nigeria. This example elucidates the essence of tailoring mobile learning solutions to meet the specific needs and challenges inherent in the local educational sector. Such an approach could be instrumental in devising mobile learning platforms that resonate with Afghanistan's unique cultural, linguistic, and educational context.

Lastly, the aspect of capacity building emerges as a consequential benefit of mobile learning. As illustrated in the Nigerian case, the collaborative and social facets of mobile learning can significantly contribute to capacity building among both educators and students. This, in turn, can foster a conducive learning environment, empowering educators with innovative teaching methodologies and students with enhanced learning experiences. Through the lens of these real-world case studies and scenarios, Afghanistan could glean insightful lessons, steering its educational landscape towards a more resilient, inclusive, and adaptable framework via the integration of mobile learning platforms.

# 6 Bridging Educational Gaps in Afghanistan: A Vision Forward

The integration of mobile learning platforms within Afghanistan's educational framework holds the promise of bridging significant educational gaps, fostering a culture of continuous learning, and catalyzing socio-economic development. This section explores the potential benefits, challenges, and long-term socio-economic implications of harnessing mobile learning platforms in Afghanistan.

### 6.1 Potential Benefits

Mobile learning platforms herald a significant stride towards ameliorating educational accessibility, acting as a bridge over geographical, social, and infrastructural chasms that often impede the educational journey. By virtue of their flexible and convenient modus operandi, these platforms extend an open invitation to learners, granting them the liberty to delve into educational resources at their leisure, anytime and anywhere.

A spotlight on gender inclusivity illuminates another promising facet of mobile learning platforms. The advent of remote learning, courtesy of these platforms, paints a hopeful picture for women and girls, offering them a culturally sensitive and secure conduit to pursue education. This could potentially usher in a new era where education is an accessible dream for every gender in every corner of a society.

On the cusp of technological innovation, mobile learning platforms are embracing Artificial Intelligence to usher in a wave of personalized learning. This embrace unfolds a learning landscape where the educational experience is tailored to the unique needs and learning pace of each student, making learning a more engaging and fruitful endeavor.

The ripple effect of mobile learning platforms extends to the realm of teacher training and professional development. As valuable reservoirs of resources, these platforms offer a fertile ground for educators to hone their skills, thereby elevating the quality of education imparted.

Community engagement finds a strong ally in mobile learning platforms. By birthing collaborative spaces for learning and discussions, these platforms sow the seeds for community-driven educational projects. The ensuing engagement fosters a vibrant educational ecosystem, where learning is a collective pursuit, enriched by the diverse experiences and knowledge of the community. Through these myriad avenues, mobile learning platforms are not just transforming the educational landscape, but are also weaving a tapestry of inclusive, personalized, and community-centric education, hinting at a promising horizon for the academia.

## 6.2 Challenges

The implementation of mobile learning platforms in Afghanistan faces a tapestry of challenges, each weaving into the broader narrative of educational transformation. At the forefront is the lack of robust technological infrastructure and internet connectivity. Many regions within the country are bereft of the basic digital scaffolding required to support mobile learning, rendering this promising educational frontier a distant reality for many.

Complementing the infrastructural hurdles is the issue of digital literacy. A palpable deficit in digital savvy among teachers, students, and the broader community casts a long shadow over the potential efficacy of mobile learning platforms. The limited digital literacy could stymie the effective utilization of these platforms, thereby diluting the transformative impact envisioned.

Equally pivotal is the matter of content development. For mobile learning to resonate with Afghan learners, the educational content must be culturally relevant and linguistically appropriate. The ethos, traditions, and linguistic nuances of Afghanistan need to be intricately woven into the fabric of the mobile learning content. This entails a concerted effort in curating educational material that not only educates but also engages the learners in a culturally coherent manner.

Lastly, the policy framework emerges as a linchpin in this endeavor. Establishing a supportive policy framework that aligns with national educational goals is quintessential for the sustainable integration of mobile learning platforms. Policies that foster a conducive environment for mobile learning, while also ensuring alignment with the broader educational objectives, will be instrumental in paving a sustainable pathway for the integration of mobile learning platforms in Afghanistan's educational tapestry. This multi-faceted approach, addressing infrastructure, digital literacy, content development, and policy framework, is indispensable for nurturing a fertile ground where mobile learning platforms can flourish and catalyze the educational transformation that Afghanistan envisages.

## 5.3 Long-term Socio-economic Benefits

Improved literacy rates are one of the significant advantages that mobile learning platforms can bring to the fore, especially in settings where educational accessibility and quality have been longstanding issues. By leveraging mobile technology to transcend geographical and infrastructural barriers, there's a tangible opportunity to uplift literacy rates. This enhancement in literacy, in turn, can act as a catalyst for socio-economic development, creating a ripple effect that goes beyond the individual to communities and the nation at large.

Furthermore, mobile learning platforms harbor the potential to significantly bolster workforce development. They can serve as conduits for vocational training and skill acquisition, essentially preparing individuals for the demands of the job market. By equipping individuals with requisite skills, mobile learning platforms lay the groundwork for gainful employment and even entrepreneurial ventures, thereby contributing to a robust and skilled workforce.

The narrative of community development is also intertwined with the proliferation of mobile learning platforms. As literacy rates improve and workforce development is accentuated, communities stand to benefit. The ripple effects of education seep into communities, fostering a culture of lifelong learning, and civic engagement. Individuals become more informed, engaged, and proactive community members, leading to a more harmonious and progressive community dynamic.

Lastly, the trajectory of economic growth is significantly influenced by the level of education and skill within the workforce. A well-educated workforce is synonymous with enhanced productivity, innovation, and a conducive environment for investments. The infusion of mobile learning platforms can play a pivotal role in nurturing a well-educated and skilled workforce, which in turn, is likely to attract investments, foster innovation, and spur economic growth. Through this lens, the integration of mobile learning platforms transcends the realm of education, impacting the broader socio-economic landscape, and propelling the nation towards a trajectory of sustainable growth and development.

# 7 Recommendations

The successful implementation of mobile learning platforms in Afghanistan hinges on a multi-pronged approach encompassing policy formulation, infrastructure development, and capacity-building measures. Below are the recommendations delineated along these dimensions:

- **Policy Measures**
  Policy formulation is a pivotal step towards the integration of mobile learning platforms within Afghanistan's educational framework. It's imperative that these policies are meticulously crafted to align with national educational goals and international best practices, which will not only foster a conducive environment for digital education but also ensure a standardized approach towards mobile learning integration (Traxler & Kukulska-Hulme, 2016). Moreover, incentivizing digital education is a pragmatic strategy to accelerate the adoption of mobile learning platforms. By creating a range of incentives for schools, educators, and students, a conducive and encouraging environment for digital learning can be established, thus propelling the nation towards an educational renaissance. Additionally, forging partnerships with international educational organizations can significantly augment the technical expertise, financial support, and global best practices available for nurturing mobile learning in Afghanistan. These collaborative endeavors can serve as a linchpin for bolstering the efficacy and reach of mobile learning platforms, thereby bridging the educational divide and fostering a culture of continuous learning and development.

- **Infrastructure Measures**
  Investing in the development of robust technological infrastructure is a pivotal step towards fostering a conducive environment for mobile learning in Afghanistan. Enhanced internet connectivity is at the core of this initiative, facilitating seamless deployment and accessibility of mobile learning platforms (ADB, 2012). Parallelly, promoting mobile device accessibility among students and educators is quintessential. This can be potentially realized through subsidized schemes or community-shared resources, thereby mitigating the digital divide and fostering inclusivity in digital education. Furthermore, embarking on a concerted effort towards content development is crucial. This involves curating culturally and linguistically appropriate educational content for mobile learning platforms, thereby ensuring relevance and engagement in the learning process. The intertwining of technological infrastructure, device accessibility, and pertinent content development forms a triad of strategic imperatives that are instrumental in leveraging mobile learning as a catalyst for educational transformation in Afghanistan.

- **Capacity-Building Measures**
  Conducting digital literacy training programs is a pivotal step towards the efficacious utilization of mobile learning platforms, targeting not only educators and students but extending to the broader community as well (Ally & Samaka, 2013). In tandem, offering continuous professional development opportunities for educators is paramount to ensure they are adeptly equipped to integrate mobile learning into their teaching practices seamlessly. Alongside, launching community engagement initiatives is instrumental in cultivating a conducive environment for mobile learning. Such initiatives aim at promoting collective participation and garnering robust support from the community, thereby creating a fertile ground for the successful implementation and acceptance of mobile learning solutions.

# 8 Opportunities for Collaboration

The journey towards integrating mobile learning solutions in Afghanistan's educational landscape necessitates a synergistic collaboration among diverse stakeholders. This section explores the multifaceted collaboration opportunities among academia, government, the private sector, international organizations, and non-governmental organizations (NGOs). Creating platforms that foster a collaborative ethos among local stakeholders, international organizations, and tech companies can significantly drive the mobile learning agenda forward in Afghanistan. Leveraging international aid and funding opportunities can provide the much-needed impetus for the development and deployment of mobile learning platforms, ensuring sustainability and broader impact. Furthermore, engaging in global knowledge exchange initiatives can offer invaluable insights from successful mobile learning implementations in other countries, allowing for the adaptation of best practices to the Afghan context.

## 8.1 Academia, Government, and Private Sector Synergy

Engaging academia and the private sector in dialogues with governmental bodies for co-creating supportive policies is crucial for fostering the integration of mobile learning. This collaborative approach in policy formulation can set a conducive regulatory framework that encourages the uptake of mobile learning platforms. Concurrently, fostering joint research and development initiatives can drive innovation in mobile learning technologies. These collaborative endeavors, ensuring contextual relevance and effectiveness, are instrumental in tailoring mobile learning solutions to meet the unique educational needs and challenges in Afghanistan.

Moreover, embarking on training programs aimed at enhancing digital literacy and pedagogical competencies among educators is pivotal. These capacity-building efforts prepare educators for the mobile learning paradigm, equipping them with the necessary skills and knowledge to effectively leverage mobile learning platforms in enhancing educational delivery.

Furthermore, leveraging the expertise and resources of the private sector can significantly expedite the development of the requisite technological infrastructure. This includes enhancing internet connectivity and device accessibility, which are fundamental for the successful implementation and scalability of mobile learning solutions. Through a holistic approach encompassing policy formulation, research and innovation, capacity building, and infrastructure enhancement, a conducive ecosystem for the integration and sustainability of mobile learning in Afghanistan can be fostered.

### 8.2 Global Partnership Horizon

Technical and Financial Support should be bolstered through the cultivation of partnerships with international organizations and NGOs. These partnerships are instrumental in harnessing technical expertise, financial support, and assimilating global best practices in mobile learning, as emphasized by Traxler & Kukulska-Hulme (2016). Following this, Program Deployment is a crucial step that necessitates engagement with international tech companies and NGOs. The objective is to meticulously design, deploy, and scale mobile learning programs while ensuring quality and contextual relevance are at the forefront. Impact Assessment is another critical facet, where collaboration with international bodies is essential to establish robust monitoring and evaluation frameworks. These frameworks are pivotal in ensuring the effectiveness and continual improvement of mobile learning initiatives. Lastly, Culturally Relevant Content Creation is an area where international collaborations can significantly contribute. By forging ties with international stakeholders, there's an opportunity to develop culturally and linguistically tailored educational content. This approach will notably enhance the engagement and relevance of mobile learning platforms, making them more accessible and impactful for the learners in Afghanistan.

## 9 Discussion

The narrative surrounding the assimilation of mobile learning platforms within Afghanistan's educational framework unveils a kaleidoscope of potentialities and hurdles. This segment ventures into a detailed discourse regarding the contextual implications, global viewpoints, and the forward trajectory as outlined through the article.

Upon delving into the contextual implications, three pivotal aspects come to the fore. Firstly, the promise of augmented accessibility and gender inclusivity through mobile learning platforms resonates profoundly within Afghanistan's milieu, a realm where geographical and socio-cultural barricades frequently thwart educational access, notably for girls and women. Secondly, the notion of surmounting infrastructural hindrances through mobile learning emerges as a plausible remedy to some of the ingrained educational quandaries Afghanistan grapples with. Lastly, the stress on concocting culturally and linguistically pertinent educational content accentuates the significance of a localized strategy in guaranteeing the efficacy and acceptance of mobile learning platforms.

Transitioning to a global perspective, the discourse uncovers a rich tapestry of collaborative vistas. The worldwide tableau of mobile learning discloses a myriad of collaborative opportunities. Engaging with international organizations, tech conglomerates, and NGOs can markedly augment the technical, financial, and strategic resources at the disposal for the execution of mobile learning initiatives in Afghanistan (Traxler & Kukulska-Hulme, 2016). Furthermore, the scrutiny of global best practices and real-world case studies in mobile learning illuminates the potential conduits for tailoring successful models to Afghanistan's distinctive context. Additionally, the alignment of mobile learning initiatives with global educational norms and frameworks can nurture a conducive ambiance for international collaborations and backing.

Pivoting towards the forward pathway, a triad of critical facets emerge. The trio of policy formulation, infrastructure amplification, and capacity building crystallizes as a cornerstone for the triumphant integration of mobile learning platforms in Afghanistan. Moreover, the envisaged collaboration among academia, government, the private sector, and international entities exemplifies a holistic approach towards navigating the intricacies inherent in mobile learning integration. Lastly, the potential long-term socio-economic boons of mobile learning, encompassing enhanced literacy rates, workforce development, and community engagement, underscore the broader impact of educational metamorphosis through mobile learning. This layered exploration enhances the discourse, offering a deeper comprehension of the contextual dynamics and the strategic imperatives pivotal for steering Afghanistan towards an educational renaissance catalyzed by mobile learning platforms.

## 10 Conclusion

The discourse traversed through this article illuminates the transformative potential embedded in the confluence of mobile learning platforms and Afghanistan's educational panorama. Amidst a milieu marked by infrastructural constraints and socio-political turbulence, mobile learning emerges as a beacon of hope, capable of bridging the pervasive educational gaps while fostering an ethos of inclusivity and continuous learning.

The core tenets explored encompass a profound understanding of the prevailing educational challenges in Afghanistan, underlined by infrastructural paucity, gender disparities, and the aftermath of conflict. In concert, the article delves into the technological underpinnings powering mobile learning platforms, shedding light on how Artificial Intelligence, Augmented Reality, Virtual Reality, and Cloud Computing can be harnessed to transcend the infrastructural barriers and catalyze an enhanced learning experience.

Enriched by real-world case studies, the narrative brings forth a spectrum of possibilities, illuminating the pathways through which mobile learning has transcended barriers in different contexts. The lessons gleaned therein offer a fertile ground for envisaging a tailored implementation of mobile learning platforms in Afghanistan, resonating with the unique socio-cultural and infrastructural contours of the nation.

The article further explores the multifaceted collaboration avenues among academia, government, the private sector, and international bodies. It underscores the quintessence of a synergistic approach in navigating the complex terrain of mobile learning integration, fostering a conducive ecosystem for educational transformation.

Lastly, the recommendations delineated provide a pragmatic blueprint for action, emphasizing the necessity of a robust policy framework, infrastructural enhancement, and capacity-building initiatives. The envisioned collaboration, undergirded by a global partnership horizon, portends a promising trajectory towards a holistic and sustainable impact on Afghanistan's educational framework.

In summation, the integration of mobile learning platforms is not merely a pedagogical shift; it is a paradigmatic transition towards a more accessible, inclusive, and resilient educational landscape in Afghanistan. The ripple effects of this transition have the potential to extend beyond the realms of education, contributing significantly to workforce development, community engagement, and the broader socio-economic development of the nation. Through a meticulous amalgamation of technology, policy, and collaborative endeavors, Afghanistan stands on the cusp of an educational renaissance, with mobile learning platforms serving as a pivotal conduit towards this envisioned horizon.

# References

Bureau of South and Central Asian Affairs. (2011). Background Note: Afghanistan. U.S. Department of State.

Razia, S. (2009). The Struggle for Education in Afghanistan. Journal of Development and Social Transformation, 6, 11-13.

UNICEF. (2011). Afghanistan: Annual Report. United Nations Children's Fund.

UNESCO. (2022). Protecting Education in Afghanistan. United Nations Educational, Scientific and Cultural Organization.

UNICEF. (2022). Education in Afghanistan. United Nations Children's Fund.

Global Partnership for Education (GPE). (2019). Education in Fragile and Conflict Affected Countries. GPE.

Statista. (2023). Global Mobile Learning Market Size. Retrieved from Statista Website.

Ally, M. (2009). Mobile Learning: Transforming the Delivery of Education and Training. AU Press.

UNESCO. (2017). Education and Literacy in Afghanistan. Retrieved from UNESCO Website.

Huang, R.H., Liu, D.J., Tlili, A., Yang, J.F., Wang, H.H. (2020). Handbook on Facilitating Flexible Learning During Educational Disruption. Beijing: Smart Learning Institute of Beijing Normal University.

Kukulska-Hulme, A. (2012). Mobile Learning as a Catalyst for Change. Open Learning: The Journal of Open, Distance and e-Learning, 27(3), 181-185.

Bishop, J. L., & Verleger, M. A. (2013). The flipped classroom: A survey of the research. In ASEE National Conference Proceedings, Atlanta, GA.

Driscoll, M. (2002). Blended Learning: Let's Get Beyond the Hype. IBM Global Services.

Knowles, M. (1975). Self-directed learning. Chicago: Follett.

Papert, S. (1991). Situating Constructionism. In I. Harel & S. Papert (Eds.), Constructionism (pp. 1-11). Ablex Publishing Corporation.

Siemens, G. (2005). Connectivism: A learning theory for the digital age. International Journal of Instructional Technology and Distance Learning, 2(1), 3-10.

UNESCO. (2012). Policy guidelines for mobile learning. UNESCO.

UNESCO. (2017). Transforming education through mobile learning in Brazil. UNESCO.

Vosloo, S. (2013). Mobile learning and policies: Key issues to consider. UNESCO Policy Paper 6.

West, M. (2017). Mobile Learning: Transforming Education, Engaging Students, and Improving Outcomes. Brookings Institution.

Shah, T., Mahmood, Z., & Riaz, M. T. (2019). Mobile Learning in Higher Education: Unveiling the Reality with Special Reference to Pakistan. Education and Information Technologies, 24(6), 3431-3451.

Johnson, L., Adams Becker, S., Estrada, V., & Freeman, A. (2015). NMC Horizon Report: 2015 Higher Education Edition. The New Media Consortium.

Al-Fraihat, D., Joy, M., & Sinclair, J. (2020). Evaluating E-Learning Systems Success: An Empirical Study. Computers in Human Behavior, 102, 67-86.

Huang, R., Liu, D., Tlili, A., Yang, J., & Wang, H. (2020). Handbook on Mobile Learning in Higher Education. Springer.

Kizilcec, R. F., Chen, M., Jasińska, K. K., Madaio, M., & Ogan, A. (2021). Mobile Learning During School Disruptions in Sub-Saharan Africa. AERA Open, 7, 23328584211014860. https://doi.org/10.1177/23328584211014860

Traxler, J., & Kukulska-Hulme, A. (2005). Mobile Learning in Developing Countries. http://hdl.handle.net/11599/77

UNESCO. (2023). Best Practices in Mobile Learning. UNESCO. Retrieved from https://www.unesco.org/en/digital-education/mobile-learning-practices#:~:text=Best%20Practices%20in%20Mobile%20Learning%2C,The%20project

Traxler, J., & Kukulska-Hulme, A. (2016). Mobile Learning: The Next Generation. Routledge.

Ally, M., & Samaka, M. (2013). Open Education Resources and Mobile Technology to Narrow the Learning Divide. International Review of Research in Open and Distance Learning, 14(2), 14-27.

ADB (Asian Development Bank). (2012). Information and Communication Technology for Education in Asia and the Pacific: Regional Comparative Report. ADB.